# CONTENT BASED VIDEO RETRIEVAL SYSTEMS

B V Patel[1] and B B Meshram[2]

[1] Shah & Anchor Kutchhi Polytechnic, Chembur, Mumbai, INDIA

patelbv@acm.org

[2]Computer Technology Department, Veermata Jijabai Technological Institute, Matunga, Mumbai, INDIA

bbmeshram@vjti.org.in

## ABSTRACT

*With the development of multimedia data types and available bandwidth there is huge demand of video retrieval systems, as users shift from text based retrieval systems to content based retrieval systems. Selection of extracted features play an important role in content based video retrieval regardless of video attributes being under consideration. These features are intended for selecting, indexing and ranking according to their potential interest to the user. Good features selection also allows the time and space costs of the retrieval process to be reduced. This survey reviews the interesting features that can be extracted from video data for indexing and retrieval along with similarity measurement methods. We also identify present research issues in area of content based video retrieval systems.*

## 1. INTRODUCTION

Content based Video Indexing and Retrieval (CBVIR), in the application of image retrieval problem, that is, the problem of searching for digital videos in large databases. "Content-based" means that the search will analyze the actual content of the video. The term 'Content' in this context might refer colours, shapes, textures. Without the ability to examine video content, searches must rely on images provided by the user [10].

Although the term "search engine" is often used indiscriminately to describe crawler-based search engines, human-powered directories, and everything in between, they are not all the same. Each type of "search engine" gathers and ranks listings in radically different ways.

Crawler-based search engines such as Google, compile their listings automatically. They "crawl" or "spider" the web, and people search through their listings. These listings are what make up the search engine's index or catalog. One can think of the index as a massive electronic filing cabinet containing a copy of every web page the spider finds. Because spiders scour the web on a regular basis, any changes made to a web site may affect search engine ranking. It is also important to remember that it may take a while for a spidered page to be added to the index. Until that happens, it is not available to those searching with the search engine[8].

Directories such as Open Directory depend on human editors to compile their listings. Webmasters submit an address, title, and a brief description of their site, and then editors review the submission. The hybrid search engines will typically favor one type of listing over the other however.

Video segmentation is first step towards the content based video search aiming to segment moving objects in video sequences. Video segmentation initially segments the first image frame as the image frame into some moving objects and then it tracks the evolution of the moving objects in the subsequent image frames. After segmenting objects in each image frame, these segmented objects have many applications, such as surveillance, object manipulation, scene composition, and video retrieval[10]. Video is created by taking a set of shots and composing them together using specified composition operators. Extracting structure primitives is the task of video segmentation that involves the detecting of temporal boundaries between scenes and between shots as shown in figure 1.

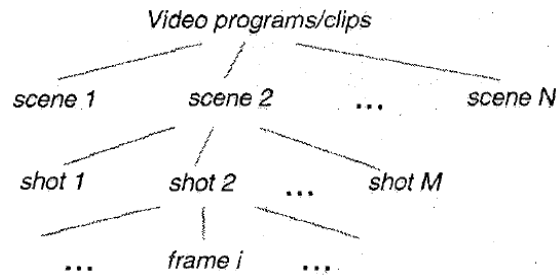

Figure 1. Video Segmentation

The first step for video-content analysis, content based video browsing and retrieval is the partitioning of a video sequence into shots. A shot is defined as an image sequence that presents continuous action which is captured from a single operation of single camera. Shots are joined together in the editing stage of video production to form the complete sequence. Shots can be effectively considered as the smallest indexing unit where no changes in scene content can be perceived and higher level concepts are often constructed by combining and analyzing the inter and intra shot relationships.

Key-frames are still images extracted from original video data that best represent the content of shots in an abstract manner. Key-frames have been frequently used to supplement the text of a video log, though they were selected manually in the past .Key-frames, if extracted properly, are a very effective visual abstract of video contents and are very useful for fast video browsing. A video summary, such as a movie preview, is a set of selected segments from a long video program that highlight the video content, and it is best suited for sequential browsing of long video programs. Apart from browsing, key-frames can also be used in representing video in retrieval video index may be constructed based on visual features of key-frames, and queries may be directed at key-frames using query by retrieval algorithms.

Once key frames are extracted next step is to extract features. The features are typically extracted off-line so that efficient computation is not a significant issue, but large collections still need a longer time to compute the features. Features of video content can be classified into low-level and high-level features.

Low-level features such as object motion, color, shape, texture, loudness, power spectrum, bandwidth, and pitch are extracted directly from video in the database. Features at this level are objectively derived from the media rather than referring to any external semantics. Features extracted at this level can answer queries such as "finding images with more than 20% distribution in blue and green color," which might retrieve several images with blue sky and green grass. Many effective approaches to low-level feature extraction have been developed for various purposes.

High-level features are also called semantic features. Features such as timbre, rhythm, instruments, and events involve different degrees of semantics contained in the media. High-level features are supposed to deal with semantic queries (e.g.,"finding a picture of water" or "searching for Mona Lisa Smile"). The latter query contains higher-degree semantics than the former. As water in images displays the homogeneous texture represented in low-level features, such a query is easier to process. To retrieve the latter query, the retrieval system requires prior knowledge that can identify that Mona Lisa is a woman, who is a specific character rather than any other woman in a painting.

The difficulty in processing high-level queries arises from external knowledge with the description of low-level features, known as the semantic gap. The retrieval process requires a translation mechanism that can convert the query of "Mona Lisa Smile" into low-level features. Two possible solutions have been proposed to minimize the semantic gap. The first is automatic metadata generation to the media. Automatic annotation still involves the semantic concept and requires different schemes for various media. The second uses relevance feedback to allow the retrieval system to learn and understand the semantic context of a query operation. Relevance feedback is discussed in section-III.

Many multimedia databases contain large numbers of features that are used to analyze and query the database. Such a feature-vector set is considered as high dimensionality. For example, Tieu & Viola used over 10,000 features of images, each describing a local pattern. High dimensionality causes the "curse of dimension" problem, where the complexity and computational cost of the query increases exponentially with the number of dimensions. Dimension reduction is a popular technique to overcome this problem and support efficient retrieval in large-scale databases. However, there is a tradeoff between the efficiency obtained through dimension reduction and the completeness obtained through the information extracted. If each data is represented by a smaller number of dimensions, the speed of retrieval is increased. However, some information may be lost. One of the most widely used techniques in multimedia retrieval is Principal Component Analysis (PCA). PCA is used to transform the original data of high dimensionality into a new coordinate system with low dimensionality by finding data with high discriminating power. The new coordinate system removes the redundant data and the new set of data may better represent the essential information.

The retrieval system typically contains two mechanisms: similarity measurement and multi-dimensional indexing. Similarity measurement is used to find the most similar objects. Multi-dimensional indexing is used to accelerate the query performance in the search process.

To measure the similarity, the general approach is to represent the data features as multi-dimensional points and then to calculate the distances between the corresponding multi-dimensional points. Selection of metrics has a direct impact on the performance of a retrieval system. Euclidean distance is the most common metric used to measure the distance between two points in multi-dimensional space. However, for some applications, Euclidean distance is not compatible with the human perceived similarity. A number of methods  have been proposed for specific purposes and presented in section IV. Rest of the paper is organized as follow: Section 2 reviews various algorithms and approaches used, section 3 we explain the various features that can be used for video indexing. Section 4 deals with the Various similarity measurement approaches, in section 5 we present research issues with content based video retrieval systems and finally we conclude in section 6.

## 2. HISTORY

Despite many research efforts, the existing low-level features are still not powerful enough to represent index frame content. Some features can achieve relatively good performance, but their feature dimensions are usually too high, or the implementation of the algorithm is difficult [3]. Feature extraction is very crucial step in retrieval system to describe the video with minimum number of descriptors. The basic visual features of index frame include color and texture [4]. Research in content based video retrieval today is a lively disciplined, expanding in breadth [5].Representative features extracted from index frames are stored in feature database and used for object-based video retrieval [6]. Texture is another important property of index frames. Various texture representations have been investigated in pattern recognition and computer vision. Texture representation methods can be classified into two categories: structural and statistical. Structural methods, including morphological operator and adjacency graph, describe texture by identifying structural primitives and their placement rules. They tend to be most effective when applied to textures that are very regular. Statistical methods, including Fourier power spectra, co-occurrence matrices, shift-invariant principal component analysis (SPCA), Tamura feature, Wold decomposition, Markov random field, fractal model, and multi-resolution filtering techniques such as Gabor and Haar wavelet transform, characterize texture by the statistical distribution of the image intensity[1].

Video Retrieval Based on Textual Queries [30] presented an approach that enables search based on the textual information present in the video. Regions of textual information are indented within the frames of the video. Video is then annotated with the textual content present in the images. Automatic Content-Based Retrieval and Semantic Classification of Video Content [31] presented a learning framework where construction of a high-level video index is visualized through the synthesis of its set of elemental features. This is done through the medium of support vector machines (SVM). The support vector machines associate each set of data points in the multi-dimensional feature space to one of the classes during training. In Content-Based TV Sports Video Retrieve Based on Audio- Visual Features and Text [32] authors propose content-based video retrieval, which is a kind of retrieval by its semantical contents. Because video data is composed of multimodal information streams such as visual, auditory and textual streams, authors describe a strategy of using multimodal analysis for automatic parsing sports video. The paper first defines the basic structure of sports video database system, and then introduces a new approach that integrates visual streams analysis, speech recognition, speech signal processing and text extraction to realize video retrieval. The experimental results for TV sports video of football games indicate that multimodal analysis is effective for video retrieval by quickly browsing tree-like video clips or inputting keywords within predefined domain.

Video Retrieval of Near–Duplicates using k–Nearest Neighbor Retrieval of Spatio–Temporal Descriptors [33] describes a novel methodology for implementing video search functions such as retrieval of near-duplicate videos and recognition of actions in surveillance video. Videos are divided into half-second clips whose stacked frames produce 3D space-time volumes of pixels. Pixel regions with consistent color and motion properties are extracted from these 3D volumes by a threshold-free hierarchical space time segmentation technique. Each region is then described by a high-dimensional point whose components represent the position, orientation and, when possible, color of the region. In the indexing phase for a video database, these points are assigned labels that specify their video clip of origin. All the labeled points for all the clips are stored into a single binary tree for efficient k-nearest neighbor retrieval. The retrieval phase uses video segments as queries. Work presented in Fast Video Retrieval via the Statistics of Motion Within the Regions-of-Interest [34] deals with very important issue to quickly retrieve semantic information from a vast multimedia database. In this work, authors propose a statistic-based algorithm to retrieve the videos that contain the requested object motion from video database. In order to speed up algorithm, authors only utilize the local motion embedded in the region-of-interest as the query to retrieve data from MPEG bitstreams. In Trajectory-Based Video Retrieval Using Dirichlet Process Mixture Models [35], present a trajectory-based video retrieval framework using Dirichlet process mixture models. The main contribution of this framework is four-fold. (1) Apply a Dirichlet process mixture model *(DPMM)* to unsupervised trajectory learning. *DPMM* is a countably infinite mixture model with its components growing by itself. (2)Employ a time-sensitive Dirichlet process mixture model *(tDPMM)* to learn trajectories' time-series characteristics. Furthermore, a novel likelihood estimation algorithm for *tDPMM* is used for the first time. (3) *tDPMM*-based probabilistic model matching scheme, which is empirically shown to be more error-tolerating and is able to deliver higher retrieval accuracy than the peer methods in the literature. (4) The framework has a good scalability and adaptability in the sense that when new cluster data are presented, the framework automatically identifies the new cluster information without having to redo the training. Theoretic analysis and experimental evaluations against the state-of-the-art methods demonstrate the promise and effectiveness of the framework. Video Annotation for Content-based Retrieval using Human Behavior Analysis and Domain Knowledge [39] presents an automatic annotation method of sports video for content-based retrieval. This approach incorporates human behavior analysis and specific domain knowledge with conventional methods, to develop integrated reasoning module for richer expressiveness of events and robust recognition.

As presented in Table- I, our objective is not exhaustive survey or completeness of research in this area, but a qualitative discussion of approaches as they evolved in this area to present our perspective on the evolution of this research field.

TABLE I: Comparison of algorithms used for feature extraction and its retrieval application

| Algorithm | Approach | Features used | Retrieval Application |
|---|---|---|---|
| Color-Texture Classification | Segment images into regions | Color, Texture Features | Perceptual importance |
| Multi-Modal Content Based Browsing and Searching Methods | Peer2Peer retrieval systems | Key Frame Extraction, Shape Feature | Object Based |
| Character Identification | Object retrieval of movie | Segmentation | Object Based |
| Semantically Meaningful Summaries | Video sequence parsing to indentiy relevant camera views, and tracks ball movements | Scripted based, Co-occurrence Matrices | Sports Video |
| Semantic Video Retrieval | Automatic Audio categorization | Audio | music, speech, etc |
| Recognizing Object in Video Sequences | Sequential classification methods for blob | Kalman filter applied over blob | Image Object based |
| Automated Scene Matching | Image matching | Wavelets, Color Descriptors | 3D scene in a movie |
| Content Based Face Retrieval | Image matching | self-organizing maps (SOMs) and user feedback Visual clustering | Face Recognition |

## 3. VIDEO INDEXING AND RETRIEVAL

Video indexing is a process of tagging videos and organizing them in an effective manner for fast access and retrieval. Automation of indexing can significantly reduce processing cost while eliminating tedious work[4]. The conventional features used in most of the existing video retrieval systems are the features such as color, texture, shape, motion, object, face, audio, genre etc. It is obvious that more the number of features used to represent the data, better the retrieval accuracy. However, as the feature vector dimension increases with increasing number of features, there is a trade off between the retrieval accuracy and complexity. So it is essential to have minimal features representing the videos, compactly. In this paper we discuss Video Key Frame Indexing, Texture, Color, Shape, audio features, etc used for indexing and retrieval.

### 3.1. Texture Features

Texture can be defined as the visual patterns that have properties of homogeneity that do not result from the presence of only a single color or intensity.

Tamura et al (1978) proposed a texture feature extraction and description method based on psychological studies of human perceptions. The method consists of six statistical features, including coarseness , contrast, directionality, line-likeness, regularity and roughness, to describe various texture properties.

Gray co-occurrence matrix (GLC) is one of most elementary and important methods for texture feature extraction and description. Its original idea is first proposed in Julesz (1975). Julesz found through his famous experiments on human visual perception of texture, that for a large class of textures no texture pair can be discriminated if they agree in their second-order statistics. Quantized index frame with GLC matrix is shown in figure 2.

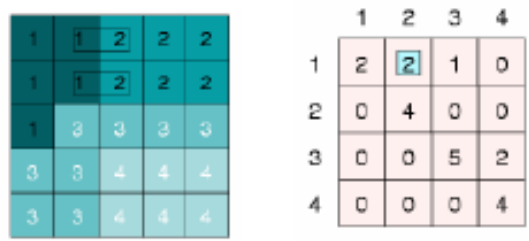

Figure 2. Quantized fame on left  with GLC on right

Following Index frame textures are extracted from GLC matrix.

$$\text{Entropy} = - \sum_i \sum_j C(i, j) \log C(i, j) \tag{1}$$

$$\text{Energy} = \sum_i \sum_j C(i, j)^2 \tag{2}$$

$$\text{Homogeneity} = -\sum_i \sum_j C(i, j) / (1 + |i{+}j|) \tag{3}$$

Contrast measures how grey levels q; q = 0, 1, ..., qmax, vary in the image g and to what extent their distribution is biased to black or white. The second-order and normalized fourth-order central moments of the grey level histogram (empirical probability distribution), that is, the

$$F_{con} = \sigma / \alpha_4^n$$

variance, $\sigma 2$, and kurtosis, $\alpha 4$, are used to define the contrast

where

$$\alpha_4 = \frac{\mu_4}{\sigma^4}; \quad \sigma^2 = \sum_{q=0}^{q_{max}} (q-m)^2 \Pr(q \mid \mathbf{g}); \quad \mu_4 = \sum_{q=0}^{q_{max}} (q-m)^4 \Pr(q \mid \mathbf{g})$$

(4)

## 3.2 Wold Decomposition Based and Gabor Texture Features

If a texture is modeled as a sample of a 2D stationary random field, the Wold decomposition can also be used for similarity-based retrieval. In the Wold model a spatially homogeneous random field is decomposed into three mutually orthogonal components, which approximately represent periodicity, directionality, and a purely random part of the field.

A 2D Gabor function $\gamma(x,y)$ and its Fourier transform $\Gamma(u,v)$ are as follows

$$\gamma(x,y) = \frac{1}{2\pi\delta x\delta y} exp \left[ -\frac{1}{2} \left( \frac{x^2}{\delta x^2} + \frac{y^2}{\delta xy^2} \right) + 2\pi\sqrt{-1Wx} \right]$$

(5)

$$\tau(u,v) = exp \left[ -\frac{1}{2} \left( \frac{(u-W)^2}{\delta v^2} \right) + \frac{v^2}{\delta v^2} \right]$$

(6)

where $\sigma_u = 1/2\pi\sigma_x$ and $\sigma_v = 1/2\pi\sigma_y$. The Gabor function is a product of an elliptical Gaussian and a complex-plane wave and it minimises joint 2D uncertainty in both spatial and frequency domain. Appropriate dilations and rotations of this function yield a class of self-similar Gabor filters for orientation- and scale-tunable edge and line detection. The filters form a complete but non-orthogonal basis set for expanding an image and getting its localised spatial frequency description. The total number of Gabor filters is equal to the product of the numbers of scales and orientations.

## 3.3. Color Features

Color is one of the most widely used visual features in multimedia context and image / video retrieval, in particular. To support communication over the Internet, the data should compress well and be suitable for heterogeneous environment with a variety of the user platforms and viewing devices, large scatter of the user's machine power, and changing viewing conditions. The CBIR systems are not aware usually of the difference in original, encoded, and perceived colors, e.g., differences between the colorimetric and device color data.

## 3.3.1. Color Descriptors

Color descriptors of images and video can be *global* and *local*. Global descriptors specify the overall color content of the image but with no information about the spatial distribution of these colors. Local descriptors relate to particular image regions and, in conjunction with geometric properties of these latter, describe also the spatial arrangement of the colors. In particular, the MPEG-7 color descriptors consist of a number of histogram descriptors, a dominant color descriptor, and a color layout descriptor (CLD).

### 3.3.2. Color histograms

A color histogram $\mathbf{h}(image)=(h_k(image) \quad k=1,...,K)$ is a $K$-dimensional vector such that each component $h_k(image)$ represents the relative number of pixels of color $C_k$ in the image, that is, the fraction of pixels that are most similar to the corresponding color. To built the color histogram, the image colors should be transformed to an appropriate color space and quantised according to a particular codebook of the size $K$.

### 3.3.3. Color Correlogram

A color correlogram of an image is a table indexed by color pairs, where the k-th entry for (i , j) specifies the probability of finding a pixel of color at a distance from a pixel of color in the image. Such an image feature turns out to be robust in tolerating large changes in appearance of the same scene caused by changes in viewing positions, changes in the background scene, partial occlusions, camera zoom that causes radical changes in shape, etc. This feature distills the spatial correlation of colors, and is both effective and inexpensive for content-based image retrieval. The correlogram robustly tolerates large changes in appearance and shape caused by changes in viewing positions, camera zooms, etc. The color correlogram is neither an image partitioning method nor a histogram refinement method. Unlike purely local properties, such as pixel position, gradient direction, or purely global properties, such as color distribution, correlograms take into account the local color spatial correlation as well as the global distribution of this spatial correlation.

### 3.4. High-level Semantic features

Semantic Gap refers to the difference between the limited descriptive power of low-level index frame features and the richness of user semantics. To support query by high-level concepts, system should provide full support in bridging this 'semantic gap' between numerical index frame features and the richness of human semantics.

In this survey we have considered the techniques used in reducing the semantic gap into five categories which are most widely used:

### 3.4.1. Using object ontologies to define high-level concepts: For databases with specifically collected images, simple semantics derived based on object-ontology may work fine, but with large collection of images, more powerful tools are required to learn the semantics.

### 3.4.2. Using machine learning tools to associate low-level features with query concepts: The techniques mentioned are Machine Learning, Bayesian Classification, Neural Networks, etc. The disadvantages of these techniques is that they require a large collection of image database for learning the data.

### 3.4.3. Introducing relevance feedback (RF) into retrieval loop for continuous learning of users' intention: Most of the current RF-based systems uses only the low-level key frames features to estimate the ideal query parameters and do not address the 'semantic' content of the index frame.

### 3.4.4. Generating semantic template (ST) to support high-level image retrieval: This technique improves the retrieval accuracy compared to traditional methods using color histogram and texture features.

**3.4.5. Making use of both the visual content of key frames and the textual information obtained from the Web for WWW (the Web) image retrieval:** Advantage is that some additional information would help for semantic-based image retrieval.
Following table 2 summarizes various techniques related to semantic features.

Table 2. Comparison of semantic features techniques

| No. | Technique | Description | Advantage | Disadvantage |
|---|---|---|---|---|
| 1 | Object Ontologies to define high-level concepts | Semantics can be derived using low-level image features and are used for high -level query. Different techniques such as color naming are used to describe the images. | Its easier to classify images based on their characteristics like color, etc. | For databases with specifically collected images, simple semantics derived based on object-ontology may work fine, but with large collection of images, more powerful tools are required to learn the semantics. |
| 2 | Machine learning tools to associate low-level features with query concepts | Using binary Bayesian classifier, high-level concepts of natural scenes are captured from classified into general types as indoor/outdoor, and the outdoor images are further classified into city/landscape, etc. | The outcome here is based on the set of input measure and using techniques of artificial intelligence they can be more precise and help reduce the semantic gap | They require a large collection of iindex frame database for learning the data. Conventional learning algorithms suffer from two problems: (1) a large amount of labelled training samples are needed, and it is very tedious and error-prone to provide such data; (2) the training set is fixed during the learning and application stages. |
| 3 | relevance feedback (RF) | The system provides initial retrieval results through query-by-example, sketch, etc. User judges the above results as to whether and to what degree, they are relevant (positive examples)/irrelevant (negative examples) to the query. Machine learning algorithm is applied to learn the | RF is an on-line processing | Most of the current RF-based systems uses only the lowlevel iindex frame features to estimate the ideal query parameters and do not address the 'semantic' content of index frames. |

| | | user' feedback | |
|---|---|---|---|
| 4 | Generating semantic template (ST) to support high-level video retrieval | tTe user first defines the template for a specific concept by specifying the objects and their spatial and temporal constraints, the weights assigned to each feature of each object. This initial query scenario is provided to the system. Through the interaction with users, the system finally converges to a small set of exemplar queries that 'best' match (maximize the recall) the concept in the user' mind. | This technique improves the retrieval accuracy compared to traditional methods using color histogram and texture features. It is the most promising for video retrieval. | Not used widely |
| 5 | Making use of both the visual content of index frames and the textual information obtained from the Web for WWW (the Web) video retrieval | The data related to index frames is used to describe them. The URL of index frame file often has a clear hierarchical structure including some information about the index frame such as index frame category. | Some additional information on the Web is available to facilitate semantic-based index frame retrieval. | Retrieval precision is poor as they cannot confirm whether the retrieved index frames really contain the query concepts. The result is that users have to go through the entire list to find the desired video. This is a time-consuming process as the returned results always contain multiple topics which are mixed together. |

## 3.5. Audio Features

Following audio features are also being used for indexing video and retrieval of video.

### 3.5.1. Short Time Energy:

The main usage of this feature is for separating speech from non-speech segments in the audio signal. It is very useful in noisy environments, because noise signals have lower average short time energy than regular speech. The average short time energy of m samples can be expressed using next term:

$$E_m = \frac{1}{N} \sum_{n=0}^{N-1} (x(n)w(m-n))^2$$

(7)

### 3.5.2. Pitch:

Pitch (fundamental frequency – F0) is a very important feature for audio analysis, especially for detection of emphasized human speech. It represents the leading frequency of a complex audio signal. In speech, pitch gets higher values as a result of speaker excitement. We decided to use the autocorrelation function in our approach of pitch estimation. The autocorrelation function for a random signal is defined as

$$A(k) = \frac{1}{2N+1} \sum_{n=-N}^{N} x(n)x(n+k)$$

(8)

From equation stated above we can compute peak values for different values of k. Pitch for the particular window is defined as a location of the largest peak value (max(A(k)) ) in the selected window, if the autocorrelation function is above a particular threshold (0.3 in our work). If last constraint is satisfied, we can calculate pitch as:

$$P = \frac{f}{k_{max}} \quad [Hz]$$

(9)

### 3.5.3. Mel-frequency cepstral coefficients:

This is a type of phoneme-level features for characterizing audio signals. It is also based on sub band division of entire frequency spectrum. MFCC are based on Mel-scale. Mel-scale is gradually warped linear spectrum, with coarser resolution on a higher and finer resolution on lower frequencies. It aim at providing a compact representation of the spectral envelope of an audio signal. MFCC-like features, for transform based MPEG-1 and AAC are computed on long, symmetric windows to estimate long-term statistics whereas for transform based 8xMDCT they are computed on frame-by-frame basis, where vector of feature is computed for each frame of 8192 samples.

### 3.5.4. Pause rate:

The pause rate feature intends to determine the quantity of a speech in an audio clip. The pause rate can be used as an indication of emphasized human speech. It can be easily calculated by counting the number of silent audio frames in an audio clip. If we denote the number of silent segments in an audio clip u as u S , we can write:

$$S_u = \sum_{j=0}^{k} \begin{cases} 1 & \text{if silent audio frame} \\ 0 & \text{otherwise} \end{cases}$$

(10)

Then we can average this on each audio clip with the number of audio frames in a clip ( |u| ) and obtain pause rate ( Pr ):

$$P_r = \frac{S_u}{|u|}$$

(11)

### 3.5.5 Onset Detection:

Onset detection functions are mid-level representations that aim at localizing transients in an audio signal. A reference onset detection function based on Hanning analysis window with length 2048 samples, which gives time resolution of 23.2ms at 44.1 kHz is used. Onset detection function for MPEG-1 and AAC are same and time resolution is 576(13ms) and 1024(23.2ms) samples at 44.1 kHz. This function is simpler for 8xMDCT and gives time resolution with respect to reference function with 512 samples.

### 3.5.6. Chromagram:

A chromagram or pitch class profile (PCP) is generally defined as a 12-dimensional vector where each dimension corresponds to the intensity of a semitone class (chroma). A reference chromagram is based on on constant Q-transform. The algorithm used by the author for computing transfom based chromagram computation is same for the three codecs. The input to the algorithm is best frequency resolution of the codecs. Obviously, the results for 8xMDCT codec are better than the other two codecs because they have highest frequency resolution.

### 3.5.7. Latent Perceptual

A whole audio clip is represented as a single vector in a latent perceptual space (LPS). This makes the computationally intensive signal-based similarity measure manageable. The method also brings out an underlying latent perceptual structure of audio clips and measures similarity based on this. The audio database is divided into 20 mutually different categories (such as aeroplane, industry, crowd, construction). The categories are derived using the available text captions.

From an audio clip feature vector can extracted. Then it is characterized by calculating the number of feature-vectors that are quantized into each of the reference clusters of signal features. There is a sparse matrix where each row represents a quantitative characterization of a complete clip in terms of the reference clusters. The reference clusters are obtained by unsupervised clustering of the whole collection of features extracted from the clips in the database. By singular-value decomposition (SVD), this sparse representation is mapped to points in the LPS. Thus each audio clip is represented as a single vector in the perceptual space.

### 3.6. Other Features and representation

### 3.6.1. Shape Features

Statistical pattern recognition approach has been prevalent for many years for shape recognition. A set of measurements are made which independently characterize some aspect of the shape. A large collection of examples, characterize the shape statistically. Suppose, for example, the mission is to distinguish between sharks and sting rays. Measurements may include properties of a region such as area, perimeter, aspect ratio, Eigen values, convex discrepancy, and various central moments. Though the computation of such features can be challenging, we do not discuss the actual computational process here, but rather refer the reader to texts on computational geometry. Simple region growing algorithm is to segment a black-and-white image in regions.

### 3.6.2. Key-Object representation

Key-frames provide a suitable abstraction and framework for video browsing. Key object are defined as the smaller units within key frame. "key-objects" used to represent key regions that participate in distinct actions within the shot. It is extremely difficult in analysis problem because key-objects do not necessarily correspond to semantic objects, we can avoid the semantic object analysis problem by seek out regions of coherent motion. Motion coherence might capture some aspect of objects desirable in retrieval. Several attributes that can be attached to key objects include color, texture, shape, motion, and life cycle. The color and texture attributes computed with algorithms.

Each shot is represented by one or more key-frames which are further decomposed into key objects. In a motion activity descriptor provides information about likely actions within the shot. It also captures the general motions in the shot such as global motions arising from camera pan or zoom. For example, motion activity descriptor can be used to distinguish "shaky" sequences captured with a hand-held camera from professionally captured sequences.

Furthermore, there are some advantages in decomposing key-object motion into a global component and a local/object-based component. In decomposition, the key-object motion can be more easily used to reflect motion relative to the background and other key-objects in the scene. Without this distinction, the key object motion would instead represent motion relative to the image frame. Thus this decomposition provides a more meaningful and effective description for retrieval.

### 3.6.3. Scale Invariant Feature Transform (SIFT) feature

Video retrieval can also be done using SIFT feature. Video is divided into frames, and frames are divided into images. The object is separated from the image by the segmentation of the image. The segmented object is a part of image. Feature is extracted from the segmented image. This method the features are extracted by using the Scale Invariant Feature Transform (SIFT) and are used to find the key points from the images because they are invariant to image.

In this method first the video is converted into images. These images are segmented using the segmentation algorithm to get the object image. Features are retrieved from the object image using the SIFT algorithm. Feature matching is then performed on database features by Mahalanobis Distance to retrieve the video from database.

Video frame referred to as a static image, is the basic unit of video data. Frame sequence is defined as a set of frame intervals, where a frame interval [i,j] is a sequence of video frames from frame i to j. The individual frames are separated by frame lines. First step is to segment a video into elementary shots, each comprising a continuous in time and space. Video stream consists of frames, shots, scenes and sequences. Physically related frame sequences generate video shots. Then combine related scenes into sequences.

Segmentation is a collection of methods allowing interpreting spatially close parts of the image as objects. It is used to locate objects and boundaries in image. Texture Segmentation used as a description for regions into segments its two types are region based and boundary based. Region-based partitions or group regions according to common image properties such as Intensity values from original images, Textures or patterns are unique to each type of region, Spectral profiles that provide multidimensional image data. Boundary-based methods used to look for explicit or implicit boundaries between regions corresponding to different issue types.

SIFT (Scale Invariant Feature Transform) features are used in object recognition are of very high dimension. They are invariant to changes in scale, 2D translation and rotation transformations. The large computational effort associated with matching all the SIFT features for recognition tasks, limits its application to object recognition problems.

## 4. SIMILARITY MEASUREMENT

Similarity measurement plays an important role in retrieval. A query frame is given to a system which retrieves similar videos from the database. The distance metric can be termed as similarity measure, which is the key-component in Content Based Video Retrieval. In conventional retrieval, the Euclidean distances between the database and the query are calculated and used for ranking. The query frame is more similar to the database frame if the distance is smaller. If x and y are 2D feature vectors of database index frame and query frame respectively. Following table 3 summarizes popular distance measurement methods.

Table 3. Distance measurement methods

| Factors Name | Formula | Image/Index Frame | Computation Time | Family |
|---|---|---|---|---|
| Euclidean distance | $D_1 = \sqrt{\sum_i (x_i - y_i)^2}$ | Color, Black and white | Less | Minkowski family |
| Squared Chord | $d_{sc}(x,y) = \sum_{i=1}^{d} (\sqrt{x_i} - \sqrt{y_i})^2$ | Color | Average | Squared-chord family |
| Chi-Square | $D_7 = \sum_i \frac{(x_i - y_i)^2}{x_i + y_i}$ | Image texture | Reduce computation time but high computation cost | Squared L2 family or $\chi2$ family |
| Manhattan | $D_2 = \sum_i |x_i - y_i|$ | Color, Black and white | Less computation require | Minkowski family |
| Divergence | $\delta_{rs} = \sum_{i=1}^{d} \frac{(x_{ri} - x_{si})^2}{(|x_{ri}| + |x_{si}|)}$ | Medical images | More | Squared L2 family or $\chi2$ family |
| Wave Hedges | $\delta_{rs} = \sum_{i=1}^{d} \frac{(x_{ri} - x_{si})^2}{(|x_{ri}| + |x_{si}|)}$ | Color, Black and white | More | Intersection family |

## 5. RESEARCH ISSUES

Following are the measure research issues related to video indexing and retrieval.

**5.1. Query Language Design:** Query language is based on the concept of "semantic indicators" while the syntax captures the basic patterns in human perception of semantic categories. Compared with other methods in reducing the "semantic gap", query language is relatively ill-understood and deserves greater attention.

**5.2. High-dimensional indexing of index frame features:** As the size of the index frame database is increasing rapidly, relative speed is an important factor to be considered. Offline Multidimensional index frame data is necessary.

**5.3. Standard DBMS extended for video retrieval:** Making video-retrieval as a plug-in module in an existing DBMS would also provide with natural integration with features derived from other sources. An Integrated CBIR system would require the integration of content-based similarity, interaction with users, visualization of video database, database management for retrieval relevant videos, etc.

**5.4. Standard index frame test bed and performance evaluation model:** A standard index frame database with query set and corresponding performance measure model is highly in need for objective performance evaluation of CBVR systems.

**5.5. Standard index frame test bed and performance evaluation model:** A standard index frame database with query set and corresponding performance measure model is highly in need for objective performance evaluation of CBVR systems.

## 6. CONCLUSIONS

Despite the considerable progress of academic research in video retrieval, there has been relatively little impact of content based video retrieval research on commercial applications with some niche exceptions such as video segmentation. Choosing features that reflect real human interest remains an open issue. One promising approach is to use Meta learning to automatically select or combine appropriate features. Another possibility is to develop an interactive user interface based on visually interpreting the data using a selected measure to assist the selection process. Extensive experiments comparing the results of features with actual human interest could be used as another method of analysis. Since user interactions are indispensable in the determination of features, it is desirable to develop new theories, methods, and tools to facilitate the user's Involvement.

**Authors**

**B. V. Patel,** is working as Principal at Shah & Anchor Kutchhi Polytechnic, Mumbai, India. He received his B.E. in Computer Science and Engineering from Karnataka University, India, M.E. in Computer Engineering from VJTI, Mumbai, India. He has authored more than twenty papers in National and International conference and journals. He is also on editorial board of International Journals. His teaching and research interest includes Multimedia Data Mining, Network Management, Network Security, E-learning, etc. He is member of ACM, INENG, IACSIT and CSI. 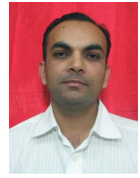

**Dr. B. B. Meshram,** is working as Professor &
Head of Computer Technology Dept., VJTI, Matunga, Mumbai. He is Ph.D. in Computer Engineering and has more than 100 papers to his credit at National and International level including international journals. He is the life member of CSI and Institute of Engineers. 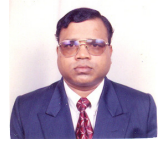